# Bigger May Not Be Better:
## An Empirical Analysis of Optimal Membership Rules in Peer-To-Peer Networks

Atip Asvanund, Karen Clay, Ramayya Krishnan, Michael Smith

H. John Heinz III School of Public Policy and Management
Carnegie Mellon University, Pittsburgh, Pennsylvania 15213
{atip, kclay, rk2x, mds}@andrew.cmu.edu

This Version: September 24, 2001

*Abstract*

Peer to peer networks will become an increasingly important distribution channel for consumer information goods and may play a role in the distribution of information within corporations. Our research analyzes optimal membership rules for these networks in light of positive and negative externalities additional users impose on the network. Using a dataset gathered from the six largest OpenNap-based networks, we find that users impose a positive network externality based on the desirability of the content they provide and a negative network externality based on demands they place on the network. Further we find that the marginal value of additional users is declining and the marginal cost is increasing in the number of current users. This suggests that multiple small networks may serve user communities more efficiently than single monolithic networks and that network operators may wish to specialize in their content and restrict membership based on capacity constraints and user content desirability.

1. Introduction

Peer to peer networks will become an increasingly important distribution channel for consumer information goods and may play a role in the distribution of information within corporations. Napster, the most prominent P2P network, was launched in 1999 as a service allowing users to share MP3 music files. In less than two years, the network's ease of use brought it 64 million registered users interested in downloading copyrighted music [CNET News.com article 1] and the attention of lawyers at the Recording Industry Association of America interested in protecting the rights of the copyright holders. A legal injunction against Napster on March 5, 2001 [Napster Injuction], as well as other publicities, such as an announcement of Napster's subscription plan on January 29, 2001 [ZDNet News Article 1], led Napster's users to migrate to a variety of alternative services including various OpenNap networks, and Gnutella. More recently, networks such as Aimster, Kazaa, and Morpheus have emerged that employ a combination of technologies derived from Napster and Gnutella.

Despite the large numbers of users and the large dollar value of the music involved, relatively little is known about peer-to-peer music sharing during the key period between the summer of 2000 and the spring of 2001 when Napster began effectively blocking trading of copyrighted material. For example, we know almost nothing about who was posting and downloading music what types of music were being downloaded, or the frequency of downloads. In the only empirical academic study of that period, Huberman and Adar found that a very small number of users were providing almost all of the content on Gnutella, the remainder were free riding on the efforts of those users.[1]

Our research analyzes optimal membership rules for these networks in light of positive and negative externalities additional users impose on the network. Network externalities would be positive if additional users provided new selections of music tracks that were previously unavailable to the network. Huberman and Adar's findings suggested that most users posted little or no music – and therefore probably no previously unavailable music – on Gnutella. Network externalities can also be negative if additional users cause congestion that makes it more difficult for existing users to find and download music. For us, this raised the question of what the optimal size might be for a music-sharing network. The answer may have seemed self-evident in the era of Napster, when a single, very large network was dominant. The coexistence of Gnutella and the Opennap networks that used the Napster protocol were not part of Napster hinted that a single large network might not have been optimal, at least not for all users. The answer to the question of the optimal network size has become relevant again in the Fall of 2001 with the rise of smaller networks.

To investigate the issue of network externalities and optimal network size, we collected data between December 19, 2000 and April 22, 2001 for the dominant networks. These networks were Napster, the OpenNap networks, and Gnutella. Data collected included on network size, song availability, login and session congestion. Data quality issues led us to focus on data from the OpenNap networks. These networks are also of interest because of their similarity to the smaller networks common in the Fall of 2001. We find that network externalities are strongly positive for song availability up to 8,000 users. Using the number of login retries as a measure of congestion, we find that the negative externalities begin to increase exponentially as the number of users on the network begin to approach the hypothesized capacity of the servers on the network. This suggests that multiple small networks may serve user communities more efficiently than single monolithic networks and that network operators may wish to specialize in their content and restrict membership based on capacity constraints and user content desirability.

---

[1] An analytical paper [Golle, Leyton-Brown & Mironov] by Golle, Leyton-Brown and Mironov used the game theoretic model to study benefits of introducing micro payment systems to P2P networks. The paper based many of its assumption on the former paper's empirical conclusion that free-riding is a problem, and it characterized how micro payments may solve free-riding and allow exchanges on P2P networks to reach equilibrium.

The rest of the paper is organized as follows. In section two we describe the data that we collected, the sources of our data, our automated data collection process, as well as the validity and advantage of our data. In section three we describe the empirical results, and how they relate to our characterization of optimal network sizes. Section four concludes, and reviews managerial implications of our findings and discusses areas for future work. It should be noted that while the focus of this paper is on network externalities that arise in the sharing of music files, the scope of P2P technology is much broader and includes applications in enterprise information sharing and in workflow applications such as collaborative engineering design. In the concluding section, we comment on the potential implications of our work for broader applications.

## 2. Data

To address characterization of optimal network sizes, we performed data collection on a number of attributes on major OpenNap networks. We used an automated software agent, which we implemented to directly observe the network by using the same protocol (i.e., the reverse engineered Napster protocol) as the client software for OpenNap networks. The data collection was performed periodically every 18 hours from December 19, 2000 to April 22, 2001. This time period ranged from the introductory stage of the OpenNap networks and lasted across the periods that the court decisions and publicities brought it to public attention, and introduced new users to the network. Observing the network longitudinally allows examination of network externalities at different network stages, which is necessary for characterizing optimal network sizes.

## 2.1 Data Collected

We tracked six major OpenNap networks, namely DJNap, FastNap, MyNapster, NakedFeet, OpenNap[2] and PowerNap, over our data collection. The data that we collected from the networks were organized to give us information on the following characteristics: 1) basic characteristics, 2) positive network externalities and 3) negative network externalities. The basic characteristics included the number of users and servers, and they are used as control variables in network externalities analysis.

The positive network externalities characteristics were measured as availability of the music tracks in our catalog of songs, and were recorded as binary variables. Our catalog of songs was generated at the beginning of our data collection. We randomly chose 5 of the 10 most popular artists for each of the 17 genres listed on Amazon.com in November 2000. We randomly chose 2 songs per artist, and in total generated the list of 170 songs. We queried every song twice to find out if it was provided by any user on the network with a broadband connection, and if it was provided by just any user regardless of connection types. This not only allowed us to analyze availability at genre levels, but also enabled us to study the effects of connection types on availability.

---

[2]There existed a sub-network in the OpenNap networks that is also called OpenNap.

There is a wide range of song availability across genres and connection types. Figure 1 depicts the variation in the average availability of each genre from our catalog for a network with 1,000 users. Availability was also differentiated for the connection types. The data was averaged across all networks and time periods of our data collection. The figure shows that on average, a network of 1,000 users had over 70% of the songs in the Pop Genre, the most available genre, available on broadband connections, but if we drop the broadband restriction, then over 80% of the songs became available. On the other hand, 0.8% of the songs in the Emerging Artists Genre, the least available genre, were available on broadband connections. By dropping the broadband restriction, 1.2% of the songs in the genre became available. We saw enough evidence of variations in the availbility across the genres and connection types to incorporate them as control variables in our model.

The negative network externalities characteristics, on the other hand, were collected as login and session congestion. Login congestion was measured by the number of retries taken to login to the network, and session congestion was measured as the average length of time taken to search for each of the 170 songs in our catalog. The congestion variables collected allows us to perform analysis on network externalities by using the basic characteristics as control variables. The variables collected for our data collection are summarized in Table 1 and selected descriptive statistics are provided in Table 2.

## 2.2 Background

The OpenNap networks were created as a non-commercial alternative to Napster. An opensource group[3] reverse-engineered Napster's technology, and used its protocol to implement the supporting infrastructures. OpenNap can be accessed by using Napigator software[4], which lists its available networks. After choosing the OpenNap network to join, users can use existing Napster software to operate on the network, because OpenNap is based on Napster's protocol.

---

[3] http://opennap.sourceforge.net

As with Napster, every OpenNap network has one or more central servers managing the central database, which maintains the listing of the music tracks that are available on the network. The central database processes all search queries and responds by giving references to the location (i.e., the IP address of the user's machine) from which the requested tracks can be downloaded. To participate in an OpenNap network, users log in to any server of the network. The search domain of a network covers all the music tracks provided by the users who are connected to any servers of that network. In other words, a user can only search the repository of the network they are connected to and cannot search the repositories of all the OpenNap networks simultaneously.

Our use of a catalog enables availability analysis at the genres level, but it may cause our analysis to be biased to the breadth of our catalog. To alleviate the issue, we resolved to generate the catalog to cover all music genres on Amazon.com, so that our catalog would cover a wide range of songs. We omitted classical music from our genres because of its unique naming convention. Searches for classical music on P2P networks cannot be uniformly automated, as with other genres, and therefore are not accurate. In addition to the breadth problem, our use of songs from the best selling artists of their respective genres may bias our study towards popular songs, but using a larger catalog would make our study less manageable.

We used the same songs in our catalog throughout our data collection, and this may cause some songs to age and become less popular, reducing availability. Our songs were randomly chosen from the list of best selling artists, and not the list of best selling songs. Best selling artists maintain popularity longer than best selling songs, as they are ranked by their overall sales across a longer period of time. For all genres with exception for Emerging Artist, whose section was removed from Amazon.com listing, the list of best selling artists did not change over our data collection. In addition, most songs in our catalog were not new songs when we added them to the catalog, but were the long-standing popular songs. Nevertheless, we included dummy variables to control for any deteriorating effect of time on these songs.

---

[4] http://www.napigator.com

An important part of our data analysis is with respect to login congestion. The user count data that we have may not reflect true user count when our agent could not log in to collect it because user count is only recorded once the agent can successfully login to the network. Thus, it is possible that while we were unable not log in to the network, user count was higher than when we could finally log in. This may cause some biases in our data interpretation. But, for analyzing the optimal network sizes, we could consider the user count data we have to be the lower bound for having such login congestion, as the user count data that we can record happened just below where we would experience congestion and not be able not login.

3. Results

Our methodology uses three tests to analyze network externalities. As discussed, we expect that additional users will provide a positive externality to the network based on the content they provide and measure this using a logit regression of network availability on user count . We also expect that additional users will place a negative externality on the network through congestion. We measure this in two ways; first with a zero-inflated Poisson regression [Cameron& Trivedi] of the number of login retries on user count, and second with a Curvilinear OLS regression of log-transformed session congestion on user count. We describe the results of each of these analyses below.

## 3.1 File Availability

File availability provides the most obvious measure of positive network externalities in our setting. The (positive) value of a user to the network is primarily a function of new content they provide other network participants. While a user who brings previously unavailable material to a network may provide significant value to other participants, a user who provides the one thousandth copy of Britney Spears' "Baby One More Time" will provide almost no value — even to fans of the pop diva. Thus, we expect to find that the marginal value of new users to the network will decline rapidly with the number of existing network participants and that the marginal decline will be more rapid for more well-represented genres.

To empirically analyze this effect, Table 3 shows results for a logit regression of file availability (0/1) onto user count; broadband access; and genre, network, and time fixed effects. The coefficient on user count, the variable of interest, shows clearly that user count has a strong positive effect on the probability that a song is available on a network.

This effect can be seen more clearly in Figure 2. Figure 2 uses our estimated coefficients to simulate graphs of file availability as a function of user count. We use the "typical" responses for the network and time fixed effects as our reference point in this figure and separately graph the response to the pop, jazz, and emerging artist genres to demonstrate the effect of genre representation on availability. As expected, Figure 2 shows that the marginal benefit of new users declines with increasing user count. Further, the effect of a new user varies dramatically based on the likelihood the file is provided by other users. For all but the smallest network,

FastNap, new users supplying pop titles provide very little marginal value to the network. For the jazz genre, the marginal effect of new users is substantial until around 10,000 users. Users providing emerging artist content, the least represented genre, provide significant value over the entire range of typical network sizes.

### 3.2 Login Congestion

Login congestion is one measure of the negative externality additional users place on the network. If the number of users trying to access a network is larger than the network's capacity, some users will have to wait, and this waiting time will increase as a function of the total number of users waiting to login.

We use the zero-inflated Poisson model to measure this externality [Lambert; Cameron & Trivedi; Nagin, Land]. The Poisson model is a useful starting point for our analysis because it captures the behavior of a count dependent variable with a long tail. Zero-inflation controls for the fact that below network capacity no retries are necessary; thus two different processes generate the zero and positive outcomes observed in our data.

The specific model is as follows. The probability of having zero login retries is first derived from the standard logit model:

(1) $$P[LR(usr\_cnt, srv\_cnt, network, period) = 1] = \frac{1}{1 + e^{-M}}$$

where M is given as

(2) $$M = \alpha_0 + \alpha_{usr\_cnt} usr\_cnt + \alpha_{srv\_cnt} srv\_cnt + \sum \alpha_{network_k} network_k + \sum \alpha_{peroid_j} period_j$$

We then analyze the predictors of multiple retries using the standard Poisson distribution:

(3) $$P[LR(usr\_cnt, srv\_cnt, network, period) = y] = \frac{e^{-\lambda}\lambda^y}{y!}$$

where

(4) $$\ln \lambda = \tau_0 + \tau_{usr\_cnt} usr\_cnt + \tau_{srv\_cnt} srv\_cnt + \sum \tau_{network_k} network_k + \sum \tau_{peroid_j} period_j$$

Table 5 presents the results of the zero-inflated Poisson model where the number of login retries necessary to gain access to the network is modeled as a function of user count and server count, along with network and time period fixed effects. As expected the number of retries increases with the user count reflecting the negative externality additional users impose on under provisioned networks. However, the server count coefficient is positive and insignificant as opposed to negative as one would expect. This may indicate poor utilization of additional servers in networks that are over capacity.

One can also conceptualize the externality of marginal users on under provisioned networks in terms of a multi-channel queuing model. If we model the capacity of the network as the number of servers and the service rate, the expected number of users in each sub-network can be modeled as expected number in the system, and would be increasing with the arrival rate.

Figure 4 displays a simulation of expected queue length holding server capacity constant at 50 and increasing the arrival rate from 1 to 50. As expected, this simulation shows that login retries increase rapidly as the number of users increases above the network capacity. This effect is analogous to the results above for the zero-inflated Poisson model. This effect can also be seen in Figure 5, which plots the number of login retries for the PowerNap network as a function of the number of users. The noise in the tail is likely a result of changing server capacity, which is controlled for in the zero-inflated Poisson regression above; however, the result is still the same: additional users provide a negative externality on networks that are above capacity.

## 3.3 Session Congestion

Session congestion is a second important type of negative externality in P2P networks. When users request files, they place traffic demands on the system, potentially degrading network performance for other users. Thus, we expect that the delay in receiving an answer to file availability queries will increase with the number of users on the network.

Table 5 measures this effect using a log-linear OLS regression of the time it takes to return results for title availability queries onto user count, server count, and network and time fixed effects.[6] As expected this regression shows that session congestion is increasing in the number of users. Thus, user queries place a negative externality on the network by increasing congestion at the central file list maintained by the network.

## 4. Conclusions

Peer to peer networks will likely become an important distribution channel for consumer information goods and may play a role in the distribution of information within corporations. However, in spite of their likely importance, little is known about the optimal design of P2P networks.

---

[5] Gnutella-based networks do not have a centralized list of available files and thus the comparison between Napster-based and Gnutella-based networks would make an interesting topic for future research.
[6] We used the log transformation of the dependent variable because it provided a better fit than the untransformed regression.

This research analyzes one aspect of P2P network design: optimal membership rules for network participants. Specifically, we empirically analyze the positive and negative externalities additional users impose on P2P networks. To do this we analyze data collected from the six largest OpenNap-based networks. This data includes the number of users logged in to each network, the availability of a random selection of popular titles, the number of login retries necessary to gain access to each network, and the session congestion in each network.

We find that additional users provide positive externalities on the network based on the amount and quality (measured in network demand) of the files they provide. However, the marginal value of an additional user declines with the number of existing users. We also find that additional users impose negative externalities on the system both in terms of added login times and added network congestion. Further the marginal cost of additional users is rising in the number of users for both of these measures.

These results suggest that several smaller networks may serve user communities more efficiently than a single monolithic network. Specifically, for each network we posit that, because of the declining value and increasing cost of marginal users, P2P networks will have finite optimal size. Given bounded network size, network operators may wish to place restrictions on which new users will be allowed into a network (and on which existing users will be allowed to remain in the network). Based on our analysis, effective membership rules will focus on the amount and desirability of content provided by the user, the size and frequency of downloads initiated by the user, and the current size of the network in terms of users and capacity. For existing music networks, the importance of desirability of content for existing users combined with network size restrictions suggests that network operators may wish to abandon generalist content strategies, in favor of focusing on particular genres.

Future work should focus on extending our results to other domains such as peer-to-peer networks for information sharing within corporations. Future research may also wish to contrast how optimal network membership rules vary with macro-level design decisions embodied in Napster-based networks and the more recently popularized Gnutella, AIMster, and Kazaa P2P network variants.

**Table 1: Summary of variables collected**

| Variable | Description |
|---|---|
| time | Time of data collection |
| network | Name of the network |
| user_count | User count on the network |
| server_count | Number of servers on the network |
| login_congestion | Number of retries taken to login to the network |
| session_congestion | Average time taken to query the network for each song |
| song_availbility | Dummy variable for whether song is available on network |
| broadband_song_availability | Dummy variable for whether song is available over a broadband connection |

**Table 2: Descriptive statistics of variables collected**

| Variables | Average | Std. | Min | Max |
|---|---|---|---|---|
| user_count | 3118 | 2283 | 68 | 8618 |
| server_count | 7 | 3 | 1 | 15 |
| login_congestion | 26 | 56 | 0 | 297 |
| session_congestion | 10 | 17 | 0 | 90 |
| song_availbility | 0.54 | 0.5 | 0 | 1 |
| broadband_song_availability | 0.45 | 0.5 | 0 | 1 |

**Figure 1: Avg. Avail per 1,000 Users on Different Connection Types**

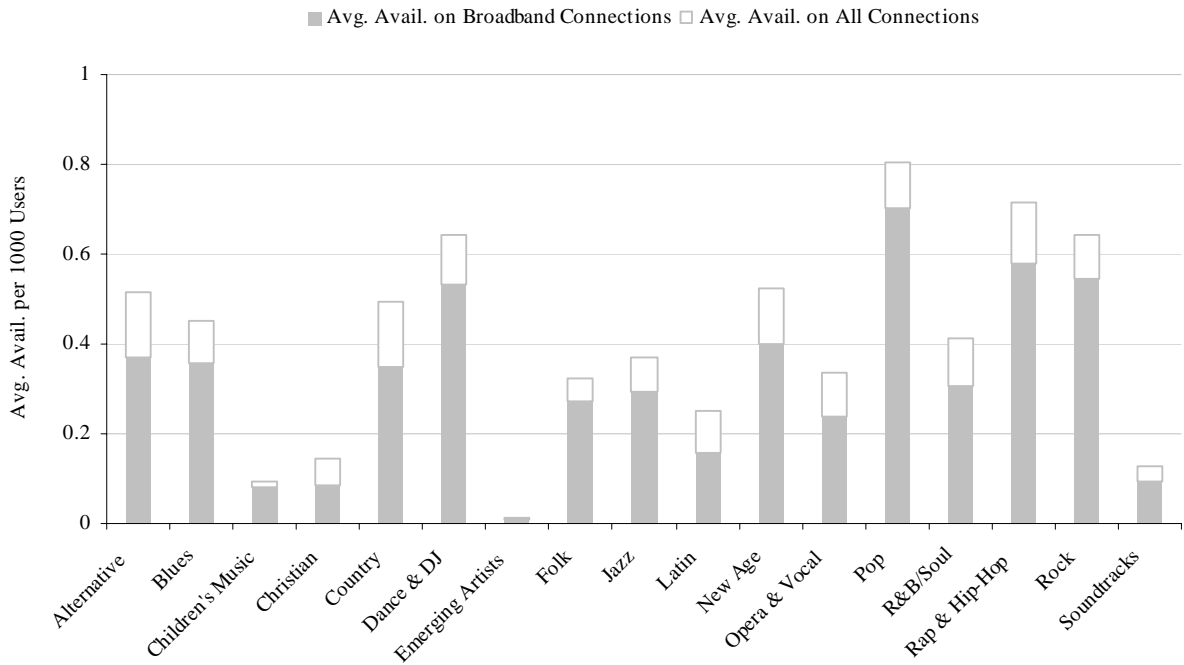

**Table 3: Logit Regression of Title Availability as a Function of User Count**

| *Availability* | *Coef.* |
|---|---|
| User Count | 7.34e-5 (7.41e-06) |
| Broadband | -.451 (.014) |
| Genre Fixed Effects | Suppressed |
| Network Fixed Effects | Suppressed |
| Time Period Fixed Effects | Suppressed |

N=109,480. Standard Errors listed in parenthesis. Fixed effect coefficients are as expected and are suppressed for simplicity.

**Figure 2: Plot of Title Availability as a Function of User Count**

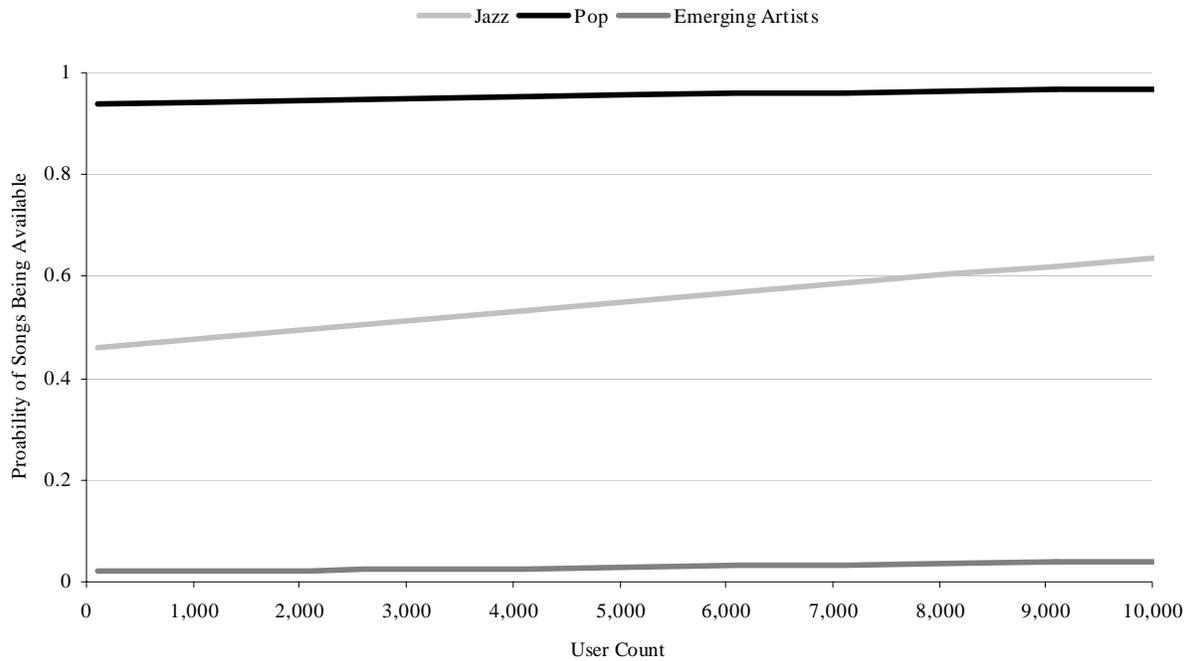

**Table 4: Zero-Inflated Regression Result**

| Login Retries | Coefficients |
|---|---|
| *Poisson* | |
| User Count | 2.86e-4 (4.87e-5) |
| Server Count | *4.48e-3 (1.46e-2)* |
| Network Fixed Effects | Suppressed |
| Time Fixed Effects | Suppressed |
| *Inflate* | |
| User Count | -7.34e-4 (1.76e-4) |
| Server Count | *-4.71e-2 (6.37e-2)* |
| Network Fixed Effects | Suppressed |
| Time Fixed Effects | Suppressed |

N=323 (128 non-zero observations). Standard Errors listed in parenthesis. Italicized coefficients are insignificant at P=.10. Fixed effect coefficients are as expected and are suppressed for simplicity.

**Figure 3: Plot of Zero-Inflated Model**

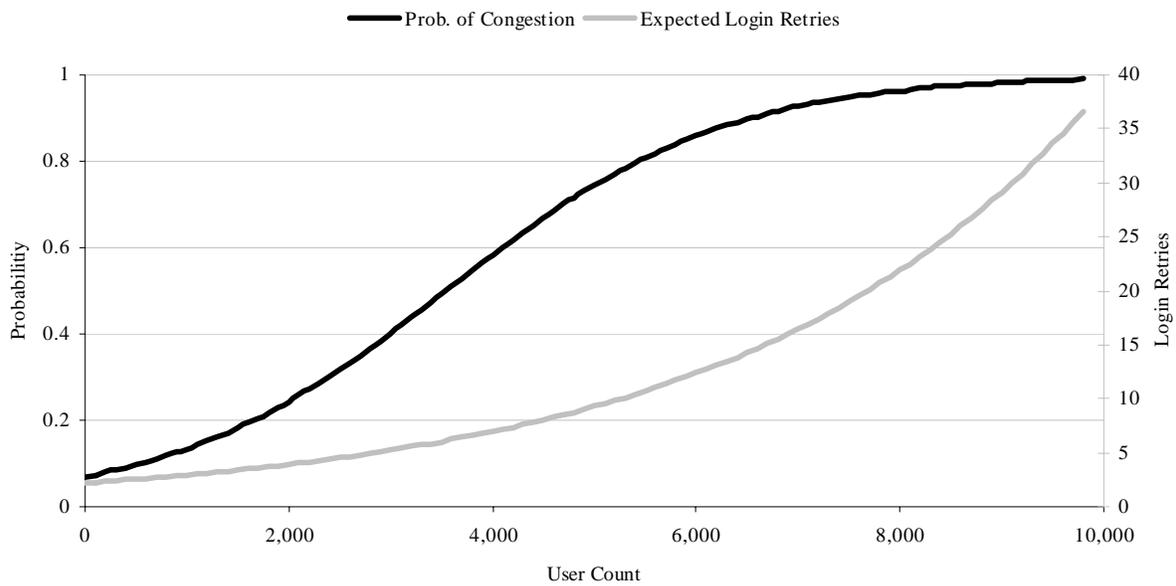

**Figure 4: Growth of Expected Queue Length Past Threshold**

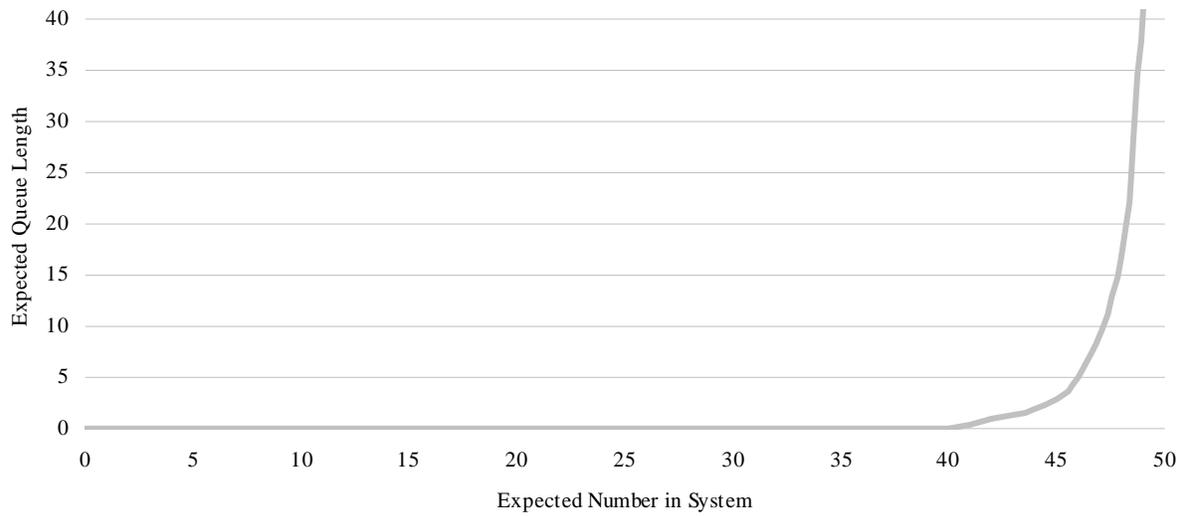

**Figure 5: Login Retries on PowerNap**

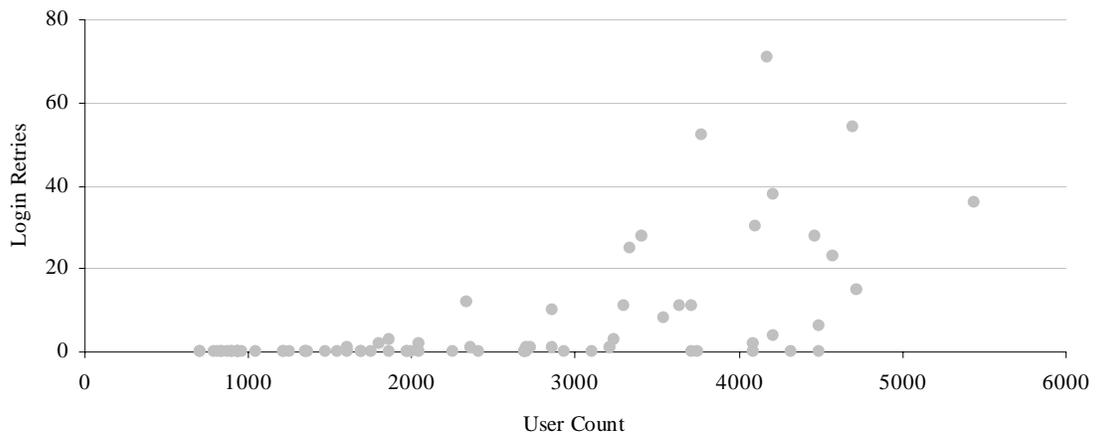

**Table 5: Session Congestion Regression**

| LN(Session Congestion) | Coefficient |
|---|---|
| User Count | 4.79e-4 (7.24e-5) |
| Server Count | -0.0847 (0.0291) |
| Network Fixed Effects | Suppressed |
| *Time Fixed Effects* | Suppressed |
| Constant | 1.531 (0.263) |

N=323. $R^2$=.465. Standard Errors listed in parenthesis. Fixed effect coefficients are as expected and are suppressed for simplicity.

**Figure 6: Session Congestion Plot**

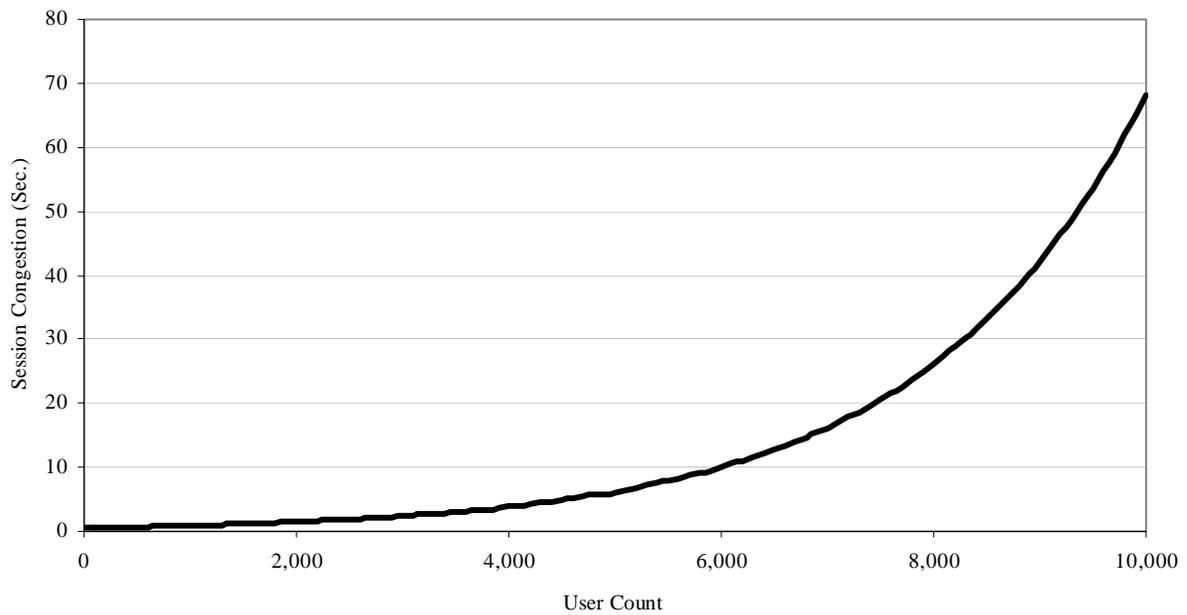